\documentclass[aps,floatfix,preprint,showpacs,preprintnumbers,nofootinbib,superscriptaddress,natbib]{revtex4}

\usepackage{graphicx,float}
\usepackage{epsfig}		
\usepackage{dcolumn}
\usepackage{bm}
\usepackage{subfigure}

\usepackage[all]{xy}
\usepackage{amsmath,upgreek}
\usepackage{amssymb}

\usepackage{pdfpages}
\usepackage{color}
\usepackage{graphicx,epstopdf}
\usepackage[colorlinks,hyperindex]{hyperref}

\def\0{\mbox{\tiny $0$}}
\def\1{\mbox{\tiny $1$}}
\def\2{\mbox{\tiny $2$}}
\def\3{\mbox{\tiny $3$}}
\def\4{\mbox{\tiny $4$}}
\def\5{\mbox{\tiny $5$}}
\def\6{\mbox{\tiny $6$}}
\def\7{\mbox{\tiny $7$}}
\def\8{\mbox{\tiny $8$}}
\def\9{\mbox{\tiny $9$}}

\def\f14{\mbox{\tiny $\frac{1}{4}$}}

\def\L{\mbox{\tiny $L$}}

\def\R{\mbox{\tiny $R$}}

\def\bm#1{\mbox{\boldmath$#1$}}

\def\bb#1{\mbox{\footnotesize $(#1)$}}

\begin{document}

\title{Global Dirac bispinor entanglement under Lorentz boosts}
\author{Victor A. S. V. Bittencourt}
\email{vbittencourt@df.ufscar.br}
\author{Alex E. Bernardini}
\email{alexeb@ufscar.br}
\affiliation{~Departamento de F\'{\i}sica, Universidade Federal de S\~ao Carlos, PO Box 676, 13565-905, S\~ao Carlos, SP, Brasil.}
\author{Massimo Blasone}
\email{blasone@sa.infn.it}
\affiliation{Dipartimento di Fisica, Universit\`a degli Studi di Salerno, Via Giovanni Paolo II, 132 84084 Fisciano, Italy}
\affiliation{Also at: INFN Sezione di Napoli, Gruppo collegato di Salerno, Italy}

\date{\today}
\renewcommand{\baselinestretch}{1.4}

\begin{abstract}
The effects of Lorentz boosts on the quantum entanglement encoded by a pair of massive spin one-half particles are described according to the Lorentz covariant structure described by Dirac bispinors. The quantum system considered incorporates four degrees of freedom -- two of them related to the bispinor intrinsic parity and other two related to the bispinor spin projection, i.e. the Dirac particle helicity. Because of the natural multipartite structure involved, the Meyer-Wallach global measure of entanglement is preliminarily used for computing global quantum correlations, while the entanglement separately encoded by spin degrees of freedom is measured through the negativity of the reduced two-particle spin-spin state. A general framework to compute the changes on quantum entanglement induced by a boost is developed, and then specialized to describe three particular anti-symmetric two-particle states. According to the obtained results, two-particle spin-spin entanglement cannot be created by the action of a Lorentz boost in a spin-spin separable anti-symmetric state. On the other hand, the maximal spin-spin entanglement encoded by anti-symmetric superpositions is degraded by Lorentz boosts driven by high-speed frame transformations. Finally, the effects of boosts on chiral states are shown to exhibit interesting invariance properties, which can only be obtained  through such a Lorentz covariant formulation of the problem.
\end{abstract}

\date{\today}
\maketitle

\section{Introduction}

Relativistic quantum information is a fast developing field that merges quantum information with relativistic quantum mechanics so to devise communication protocols in relativistic frameworks involving, for instance, clock synchronization \cite{clock}, position verification \cite{position} and teleportation protocols \cite{teleport}.
The effects of a relativistic frame transformation on quantum correlations have been recently investigated \cite{relat01, relat02, relat03, relat04, relat05, relat06, relat07, relat08, relatvedral} and, considering spin as the natural tool for quantum information engineering, the effects of frame transformations (Lorentz boosts) on the quantum entanglement encoded by a pair of spin one-half particles have been investigated.

From the kinematic point of view, the action of the linear transformation given by a Lorentz boost, $\Lambda$, describes the change of space-time coordinates from an inertial frame, $\mathcal{S}$, to another one, $\mathcal{S}^{\prime}$, which moves with respect to $\mathcal{S}$, as to set, for instance, the quadrimomenta transformation relation $p^{\nu\prime} = \Lambda^{\nu}_{\mu} p^{\mu}$, summarized by $p^{\prime} = \Lambda p$ in the matricial representative notation, where $p^{\prime}$ and $p$ are described by coordinates at $\mathcal{S}^{\prime}$ and $\mathcal{S}$, respectively.
As seminally stated by Wigner \cite{wigner}, under such a transformation between inertial frames, an observable spin (projector operator) described by the $SU(2)$ adjoint representation realized by ${\boldmath{\mbox{$\sigma$}}} = (\sigma_x,\sigma_y,\sigma_z)$, where $\sigma_{x,y,z}$ are the Pauli matrices, has its spin projection onto the particle momentum direction, $\hat{\bm{e}}_p\cdot\hat{\boldmath{\mbox{$\sigma$}}}$, with ${\bm{e}}_p={\bm{p}}/\vert{\bm{p}}\vert$, changed as to return $\hat{\bm{e}}_{p^{\prime}}\cdot\hat{\boldmath{\mbox{$\sigma$}}}  \neq  \hat{\bm{e}}_{p}\cdot\hat{\boldmath{\mbox{$\sigma$}}}$, where boldfaced variables, $\bm{v} = (v_x,v_y,v_z)$, denote spatial vectors with modulus $v = \sqrt{\bm{v} \cdot \bm{v}}$, and hats ``$~\hat{}~$'' denote quantum operators.

The rigorous treatment of the above kinematic properties, and of their imprints on quantum states of spin one-half particles, involves a description of their observable related properties in terms of the irreducible representations ({\em irreps}) of the Poincar\'{e} group \cite{fonda, weinberg}.
For instance, for a particle with momentum, $\bm{p}$, in an inertial frame $\mathcal{S}$, described by the quantum state $\vert\phi_s \bb{\bm{p}}\rangle$, where $s=1,2$ denote accessible spin states, the action of a Lorentz boost, $\Lambda$, that describes the change from $\mathcal{S}$ to $\mathcal{S}^{\prime}$, is given by the unitary transformation \cite{fonda, weinberg, wigner}
\begin{equation}
\label{spinstates}
\vert \phi_s \bb{\bm{p}}\rangle  \, \,  \rightarrow \, \, \hat{D} [\Lambda]  \vert\phi_s \bb{\bm{p}}\rangle  = \sum_{r} c_{s r}(\Lambda, \bm{p})  \vert\phi_r (\bm{p}^{\prime})\rangle,
\end{equation}
where the unitary operator, $\hat{D} [\Lambda]$, and its explicit component dependence, $c_{s r}$, according to the Poincar\'e group representations \cite{fonda, weinberg} (cf. Sec.~II), are given in agreement with the {\em irrep} of the quantum state, $\vert\phi_s \bb{\bm{p}}\rangle$, which can describe, for instance, a spinor (in a doublet representation, like electrons and positrons described either as Weyl or as Majorana fermions), a vector (in a triplet representation, like $^3S_1$ positronium, or even photons), bispinors (in a double doublet representation, like electrons and positrons described by Dirac fermions), or even scalar (in a singlet representation, like $^1S_0$ positronium) and higher order (maybe non-physical) tensor states.
The point in this paper is that when quantum states depend on the momentum, namely those described by Dirac equation solutions for spin one-half states, different inertial observers will see different superpositions, and if somehow the momentum degrees of freedom are traced out, the quantum entanglement between spin states might change \cite{relat01, relat02, relat03, relat04, relat05, relat06, relat07, relat08, relatvedral}. Of course, for two-particle states, the question related to the influence of the reference frame in the computation of quantum correlations is much more engendered in the framework of relativistic quantum mechanics supported by the Dirac formalism.

Despite the effectiveness of the {\em irreps} of the Poincar\'{e} group, in the Lorentz covariant Hamiltonian formulation of quantum mechanics, one has to pay attention to inclusion of mass in the relativistic formalism described by the Dirac equation.
As one shall see in Sec. II, it requires the inclusion of the parity symmetry and the equalization of its role with the helicity (spin one-half projection, $\hat{\bm{e}}_p\cdot\hat{\boldmath{\mbox{$\sigma$}}} \sim \hat{\sigma}_z$) symmetry, as an accomplished $SU(2)$ symmetry. It supports, for instance, the description of electrons as Dirac Hamiltonian eigenstates in the double doublet {\em irrep} of the $SU(2) \otimes SU(2)$. Spatial parity couples positive and negative parity states with positive and negative helicity states as they were described by {\em irreps} of the Poincar\'{e} group \cite{fonda} and, in order to have complete invariance under parity, one needs to consider the \textit{extended Poincar\'{e} group} \cite{weinberg,WuTung}. In this case, spin one-half is carried by Dirac four component spinors, the bispinors satisfying the Dirac equation, in a representation supported by a subgroup of $SL(2,\mathbb{C}) \otimes SL(2,\mathbb{C})$, the $SU(2) \otimes SU(2)$. 
In fact, the description of massive charged fermions (such as electrons, muons, quarks, etc...) requires the {\em irreps} of the {\em complete Lorentz group}\footnote{The complete Lorentz group is composed by the proper Lorentz group together with space inversion. The extended Poincar\'{e} group is given by the complete Lorentz group with addition of space-time translation \cite{WuTung}.}, namely the Dirac (bi)spinors \cite{WuTung}. 

The intrinsic spin-parity (or helicity-parity) entanglement exhibited by a single Dirac bispinor has already been investigated in the context of quantum correlations driven by interactions with external fields \cite{extfields}, which has been used for simulating Dirac-like systems as, for instance, four level ion traps \cite{diraclike01} and lattice/layer schemes in bilayer graphene \cite{diraclike02}. 
Generically, such an intrinsic entanglement encoded by Dirac-like $SU(2) \otimes SU(2)$ structures can also be generated, for example, by quantum electrodynamics (QED) scattering processes \cite{solano}. 

Not in the same scope, but also emphasizing the bispinor structure of fermionic quantum states, states constructed with the solutions of Dirac equation have been considered in the scrutinization of Bell inequalities \cite{bellDirac} and to obtain proper covariant spin density matrices and definitions of the position operator in the context of relativistic quantum mechanics \cite{spins, celeri}. 
Likewise, the effects of Lorentz boosts in quantum entanglement encoded in bispinors were described in connection with Wigner rotations for a specific class of states \cite{bi-spinorarxiv02}, and in the context of Fouldy-Wouthuysen (FW) spin operator \cite{greiner}, with a focus on properties of transformation of spin-spin entanglement encoded in FW eigenstates \cite{bi-spinorFW}. 
However, considering the focus on the most phenomenologically appealing measurement of two-particle spin-spin entanglement, the intrinsic $SU(2) \otimes SU(2)$ covariant structure of Dirac bispinors, which is associated with intrinsic parity and spin \cite{SU2} for each particle, has not yet been completely incorporated into such an overall relativistic framework. 

The aim of this work is therefore to estimate the influence of Lorentz boosts on the quantum entanglement encoded in the intrinsic $SU(2) \otimes SU(2)$ structure of two spin one-half Dirac particles which are also spin-spin entangled. As each particle described by Dirac bispinors carries two qubits, the whole state is a four-qubit one, and since multipartite entanglement is generally present in such states, the Meyer-Wallach global measure of entanglement \cite{globalMW} shall be considered as a measure of the entanglement encoded in the four qubits of the system. Alternatively, the net result for the spin-spin entanglement, encoded only in a two-qubit mixed state, shall be computed through the negativity \cite{negativity01, negativity02}. The effects of a Lorentz boost on the entanglement content of generic two-particle Dirac bispinor states shall be obtained for the case where superpositions of helicity plane waves are considered. The obtained results shall be specialized to anti-symmetric states showing, for example, that a Lorentz boost cannot create spin-spin entanglement in an initial separable anti-symmetric state. 

The paper is structured as follows. In Sec.~II, the complete Lorentz covariant structure of the Dirac equation solutions, namely associated to the properties of $SU(2)$ spinor doublet representations, and to the composition of higher order multiplet representations, is reported about, and the foundations for establishing and discussing the spin-parity intrinsic entanglement are introduced.
In Sec.~III, by using the intrinsic $SU(2) \otimes SU(2)$ structure of the Dirac equation, the entanglement profile of a generic superposition of Dirac bispinors is described. In Sec.~IV, by using the transformation laws of bispinors under Lorentz boosts, the effects of such transformations on the quantum correlations encoded by two-particle states are computed, with particular emphasis for anti-symmetric states. 
In addition, the investigation of the effects of boosts onto the superposition of chiral bispinors shows that some subtle invariance properties can be obtained. 
Our conclusions are given in Sec.~V, where lessons concerning the importance of accounting for the Lorentz covariant structures in the computation of quantum correlations are drawn.

\section{Lorentz covariant structure of the Dirac equation and spin-parity intrinsic entanglement}

In quantum mechanics, the free particle Dirac Hamiltonian in the coordinate space reads
\begin{equation}
\label{diracequation}
\hat{H} \, \psi\bb{x} = i \frac{\partial \, \psi\bb{x}}{\partial t} = (-i\bm{\nabla} \cdot \hat{\bm{\alpha}} + m \hat{\beta}) \,\psi\bb{x} = (-i \hat{\alpha}_i\partial^i + m \hat{\beta}) \,\psi\bb{x} = \pm E_p \,\psi\bb{x} ,
\end{equation}
where $E_p = \sqrt {p^2 + m^2}$, the space-time dependence has been resumed by $x\sim(t,\bm{x})$, and the Dirac matrix operators $ \hat{\bm{\alpha}} =(\hat{\alpha}_x,\,\hat{\alpha}_y,\,\hat{\alpha}_z)$ and $\hat{\beta}$ satisfy the anticommuting relations, $
\hat{\alpha}_i \hat{\alpha}_j + \hat{\alpha}_j \hat{\alpha}_i = 2 \delta_{ij} \hat{I}_4$, and $\hat{\alpha}_i \hat{\beta} + \hat{\beta} \hat{\alpha}_i =0$, for $i,j = x,y,z$, with $
\hat{\beta}^2 = \hat{I}_4$, $I_N$ the $N$-dim identity matrix, and $\hat{H}$ expressed in natural units, i.e. with $c = \hbar = 1$.

The above Dirac Hamiltonian dynamics exhibits some symmetries that are supported by a group representation described by a direct product between two algebras which compose a subset of the group $SL(2,\mathbb{C})\otimes SL(2,\mathbb{C})$, the group $SU(2)\otimes SU(2)$. 

To clear up this point, before discussing the above statement in the enhanced language of Lie algebra and Lie groups, one simply notices that {\em left-handed} {\em spinors} are described by a doublet ($2$-dim) representation of the $SU(2)$ (left) and a singlet ($1$-dim) representation of the $SU(2)$ (right),
$(\bm{2},\bm{1}) \equiv \psi^{\dagger} _{\L}\bb{x} =\left(
\psi _{\L\1}\bb{x},\,
\psi _{\L\2}\bb{x}\right)$ and, analogously, {\em right-handed} {\em spinors} are described by a doublet ($2$-dim) representation of the $SU(2)$ (right) and a singlet ($1$-dim) representation of the $SU(2)$ (left), $(\bm{1},\bm{2}) \equiv \psi^{\dagger} _{\R}\bb{x} =\left(
\psi _{\R\1}\bb{x},\,
\psi _{\R\2}\bb{x}\right)$,
in order to support the following decomposition for the Dirac state vectors,
$ \psi^{\dagger} \bb{x} =\left(
\psi _{\L\1}\bb{x},\,
\psi _{\L\2}\bb{x},\,
\psi _{\R\1}\bb{x},\,
\psi _{\R\2}\bb{x}\right)
\equiv (\bm{2},\bm{1})\oplus(\bm{1},\bm{2})$,
in a not unique double doublet representation of the $SU(2)\otimes SU(2)$ group.

Therefore, the free particle Dirac equation is thus mapped into coupled differential equations for {\em left-} and {\em right-handed} components, respectively,
\begin{eqnarray*}
i{\overline{\sigma}}^{\mu }\partial _{\mu }\psi _{\L}\bb{x}
-m\psi _{\R}\bb{x} &=&0, \\
i{\sigma }^{\mu }\partial _{\mu }\psi _{\R}\bb{x} -m\psi
_{\L}\bb{x} &=&0,
\end{eqnarray*}
in the so-called {\em chiral representation}, $(\hat{I}_{\2} , \hat{\bm{\boldsymbol{\sigma }}}) \equiv
\sigma^{\mu}$ and $(\hat{I}_{\2}, -\hat{\bm{\boldsymbol{\sigma }}}) \equiv {\overline{\sigma}}^{\mu}$, for which the Lagrangian density reads
\begin{equation}
\mathcal{L}=i\psi _{\L}^{\dagger }{\overline{\sigma}}^{\mu }\partial
_{\mu }\psi _{\L}+i\psi _{\R}^{\dagger }\mathbf{\sigma }^{\mu }\partial _{\mu
}\psi _{\R}-m\left( \psi _{\L}^{\dagger }\psi _{\R}+\psi _{\R}^{\dagger }\psi
_{\L}\right),
\end{equation}
from which a correspondence with the {\em spinor} chirality is identified.

As it has been mentioned, the above choice is not unique. Another particular representation of the Dirac matrices is the Pauli-Dirac representation in which
the Dirac matrices are decomposed into tensor products of Pauli matrices \cite{SU2}, as $\hat{\alpha}_i = \hat{\sigma}_x^{(P)} \otimes \hat{\sigma}_i^{(S)}$, for $i = x,y,z$ and $\hat{\beta} = \hat{\sigma}_z^{(P)} \otimes \hat{I}^{(S)}$, which has another subjacent $su(2) \oplus su(2)$ algebra from the $SU(2)\otimes SU(2)$ group which, in this case, does not correspond to {\em left-} and {\em right-handed} chiral projection representations, instead, are associated to intrinsic parity, $P$, and spin (or helicity), $S$. 
In this case, the Dirac Hamiltonian is thus re-written in terms of Kronecker products between Pauli matrices as
\begin{equation}
\label{twoqubithamiltonian}
\hat{H} = \bm{p}\cdot (\hat{\sigma}_x^{(P)} \otimes \hat{\bm{\sigma}}^{(S)}) + m ( \hat{\sigma}_z ^{(P)} \otimes \hat{I}^{(S)}),
\end{equation}
from which, according to the interpretation of quantum mechanics as an information theory for particles, where the superscripts $P$ and $S$ refer to the {\em qubits} of parity and spin, one can identify the Dirac equation solutions as they were described by two {\em qubit} states encoded in a massive particle whose dynamics is constrained by continuous variables.

Within this framework, from the Hamiltonian Eq.~(\ref{twoqubithamiltonian}), the normalized stationary eigenstates in the momentum coordinate are written in terms of a sum of direct products describing \textit{spin-parity} entangled states,
\begin{eqnarray}
\label{twoqubitspinor}
\vert \, u_s\bb{\bm{p}} \, \rangle &=& \frac{1}{\sqrt{2 E_{p} (E_{p} + m)}} \left[ (E_{p} + m)\,\, \vert + \rangle \otimes \vert \chi_s \bb{\bm{p}} \rangle \,\,+\,\, \vert - \rangle \otimes \,(\, \bm{p} \cdot \bm{\sigma} \, \vert \chi_s \bb{\bm{p}} \rangle )\right],\\
\vert \, v_s\bb{\bm{p}} \, \rangle &=& \frac{1}{\sqrt{2 E_{p} (E_{p} + m)}} \left[ (E_{p} + m)\,\, \vert - \rangle \otimes \vert \chi_s \bb{\bm{p}} \rangle\,\, + \,\,\vert + \rangle \otimes \,(\, \bm{p} \cdot \bm{\sigma} \, \vert \chi_s \bb{\bm{p}} \rangle )\right],
\end{eqnarray}
for positive and negative eigenvalues (associated frequencies), $\pm E_p = \pm\sqrt{p^2+m^2}$, respectively\footnote{In the bispinorial form, one has 
\begin{eqnarray}
\label{bi-spinors}
u_s\bb{\bm{p}} = \frac{1}{\sqrt{2 E_{p} (E_{p} + m)}}\left[ \begin{array}{rl} (E_{p} + m) &\chi_s \bb{\bm{p}}\\ \bm{p} \cdot \bm{\sigma} &\chi_s \bb{\bm{p}} \end{array}\right] \,\, \mbox{and} \,\, v_s\bb{\bm{p}} =\frac{1}{\sqrt{2 E_{p} (E_{p} + m)}} \left[ \begin{array}{rl} \bm{p} \cdot \bm{\sigma} &\chi_s \bb{\bm{p}} \\ (E_{p} + m)& \chi_s \bb{\bm{p}} \end{array}\right],
\end{eqnarray}
with the orthogonality relations identified by $u_s^\dagger\bb{\bm{p}} u_r\bb{\bm{p}} = v_s^\dagger\bb{\bm{p}} v_r\bb{\bm{p}} = \delta_{sr}$ and $u_s^\dagger\bb{\bm{p}} v_r\bb{-\bm{p}} = v_s^\dagger\bb{\bm{p}} u_r\bb{-\bm{p}} =0$, and the completeness relation given by $$\displaystyle \sum_{s=1}^2\Big[u_s\bb{\bm{p}} u_s^\dagger\bb{\bm{p}} + v_s\bb{\bm{p}}v_s^\dagger\bb{\bm{p}} \Big] =  \hat{I}_4.$$},
where $\vert \chi_s\bb{\bm{p}}\rangle \in \mathcal{H}_S$, with $s=1,\,2$, are the spinors related to the spatial motion of the particle, i.e. the particle's helicity, which describes the dynamics of a fermion (in momentum representation) coupled to its spin, and $\vert \pm \rangle \in \mathcal{H}_P$ are intrinsic parity states. 
States as described by Eqs.~(\ref{bi-spinors}) are composite quantum systems in a total Hilbert space $\mathcal{H} = \mathcal{H}_P \otimes \mathcal{H}_S$ and, in the general form of Eq.~(\ref{twoqubitspinor}), they are spin-parity entangled \cite{SU2}. 
Of course, they are superposition of orthonormal parity eigenstates, $\vert \pm \rangle $, and therefore, they do not have a defined intrinsic parity quantum number\footnote{A defined total parity operator $\hat{P}$ acts on the direct product $\left\vert \pm \right\rangle \otimes \left\vert \chi_s\bb{\bm{p}}\right\rangle$ in the form of $$\hat{P}\left( \left\vert \pm \right\rangle \otimes \left\vert \chi_s\bb{\bm{p}}\right\rangle\right) =\pm \left( \left\vert \pm \right\rangle\otimes \left\vert \chi_s(-\bm{p})\right\rangle\right),$$ and, for instance, it corresponds to the Kronecker product of two operators, $\hat{P}^{(P)}\otimes \hat{P}^{(S)}$, where $\hat{P}^{(P)}$ is the intrinsic parity (with two eigenvalues, $\hat{P}^{(P)}\left\vert \pm \right\rangle =\pm \left\vert\pm \right\rangle $) and $\hat{P}^{(S)}$ is the spatial parity (with $\hat{P}^{(S)}\chi_s \left( \bm{p}\right) =\chi_s \left( -\bm{p}\right) $). 
}.
 
To summarize, the spin degree of freedom (DoF) identified by the index ``$s$'' is associated to {\em irreps} of the proper Poincar\'{e} group, and the positive/negative associated energy eigenstates of the spin one-half particles can be re-indexed through the notation
\begin{eqnarray}
\label{twoqubitspinor22}
\vert \, u_{\pm,s}\bb{\bm{p}} \, \rangle &=& \frac{1}{\sqrt{2 E_{p} (E_{p} + m)}} \left[ (E_{p} + m)\,\, \vert \pm \rangle \otimes \vert \chi_s \bb{\bm{p}} \rangle \,\,+\,\, \vert \mp \rangle \otimes \,(\, \bm{p} \cdot \bm{\sigma} \, \vert \chi_s \bb{\bm{p}} \rangle )\right],\end{eqnarray}
for vectors belonging to the {\em irrep} labeled by $(\pm, \frac{1}{2})$, associated to the $SU(2)\otimes SU(2)$ group \cite{fonda, weinberg, wigner}. Therefore, the invariance under spatial parity symmetry requires an analysis with the {\em complete Lorentz group} in order to include {\em irreps} of $SU(2) \otimes SU(2)$ which merge spin with the additional DoF of intrinsic parity \cite{fonda, WuTung}.

In the context of a group theory, the above assertion can be better understand when the representations of $sl(2,\mathbb{C})\oplus sl(2,\mathbb{C})$, which corresponds to the Lie algebra of the $SL(2,\mathbb{C})\otimes SL(2,\mathbb{C})$ Lie group, are irreducible, i.e. they correspond to tensor products between linear complex representations of $sl(2,\mathbb{C})$, as it has been observed by considering the subgroup $SU(2)\otimes SU(2) \subset SL(2,\mathbb{C})\otimes SL(2,\mathbb{C})$. Unitary {\em irreps} of the $SU(2)\otimes SU(2)$ are built through tensor products between unitary representations of $SU(2)$, in a one-to-one correspondence with the group $SL(2,\mathbb{C})\otimes SL(2,\mathbb{C})$. Since it is a simply connected group, one also has a unique correspondence with the $sl(2,\mathbb{C})\oplus sl(2,\mathbb{C})$ algebra.

As it has been above identified for the chiral basis and for the spin-parity basis, the existence of {\em inequivalent representations} of $SU(2) \otimes SU(2)$ follows from the above mentioned one-to-one correspondences. Inequivalent representations may not correspond to all the complete set of representations of $SL(2,\mathbb{C})\otimes SL(2,\mathbb{C})$, and therefore, of the proper Lorentz transformations that compose the $SO(1,3)$ group, i.e. the Lorentz group\footnote{Instead, they describe a subset of transformations of the $SO(4) \equiv SO(3)\otimes SO(3)$ group, as for instance, those which include the double covering rotations.}.

Turning back to our point, as the transformations of $SU(2)\otimes SU(2)$ can be described by a subset of $SL(2,\mathbb{C})\otimes SL(2,\mathbb{C})$, one may choose at least two {\em inequivalent} subsets of $SU(2)$ generators, such that $SU(2)\otimes SU(2) \subset SL(2,\mathbb{C})\otimes SL(2,\mathbb{C})$, with each group transformation generator having its own {\em irrep}. Therefore, a fundamental object of the $SU_{\xi}(2)$, a spinor-like object $\xi$ described by $(\pm,\,0)$, transforms as a {\em doublet} -- the fundamental representation -- of $SU_{\xi}(2)$, and as a singlet -- a transparent object under any $SU_{\chi}(2)$ transformations.
Reciprocally, the fundamental object of the $SU_{\chi}(2)$, a typical spinor $\chi$ described by $(0,\frac{1}{2})$, transforms as a {\em doublet} of the $SU_{\chi}(2)$, and as a singlet of the $SU_{\xi}$. Under an improved notation generalized to higher dimension representations, $(\bm{dim}(SU_{\xi}(2)),\bm{dim}(SU_{\chi}(2)))$, the {\em spinor} $\xi$ is an object given by (\bm{2},\bm{1}). Following the generalized idea for an arbitrary $SU_(2)\otimes SU_(2)$ composition, one has the representations as given by

$(\bm{1},\bm{1})$ -- for {\em scalar} or {\em singlet}, with angular momentum projection $j = 0$;

$(\bm{2},\bm{1})$ -- for {\em spinor} $(\frac{1}{2},\,0)$, with angular momentum projection $j = 1/2$, which corresponds to $(\pm,\,0)$ in case of $SU_{\xi}(2)\otimes SU_{\chi}(2)$ and also applies for designating {\em left-handed} spinors in case of an inequivalent representation of the $SU_{\tiny\mbox{Left}}(2)\otimes SU_{\tiny\mbox{Right}}(2)$ group;

$(\bm{1},\bm{2})$ -- for {\em spinor} $(0,\,\frac{1}{2})$, with angular momentum projection $j = 1/2$, which also applies for designating {\em right-handed} spinors in case of an inequivalent representation of the $SU_{\tiny\mbox{Left}}(2)\otimes SU_{\tiny\mbox{Right}}(2)$ group;

$(\bm{2},\bm{2})$ -- for {\em vector}, with angular momentum projection $j = 0$ and $j = 1$; etc.

With respect to the fundamental representations of $SL(2,\mathbb{C})$, one may construct more complex objects like
$ (\bm{1},\bm{2}) \otimes (\bm{1},\bm{2}) \equiv (\bm{1},\bm{1}) \oplus (\bm{1},\bm{3})$,
a representation that composes Lorentz tensors like
\begin{equation}
C_{\alpha\beta}\bb{x} = \epsilon_{\alpha\beta} D\bb{x} + G_{\alpha\beta}\bb{x},
\end{equation}
where $D\bb{x}$ is a scalar, and $G_{\alpha\beta} = G_{\beta\alpha}$ is totally symmetric, or even
$ (\bm{2},\bm{1}) \otimes (\bm{1},\bm{2}) \equiv (\bm{2},\bm{2})$,
such that
$ (\bm{2},\bm{2}) \otimes (\bm{2},\bm{2}) \equiv (\bm{1},\bm{1}) \oplus (\bm{1},\bm{3}) \oplus (\bm{3},\bm{1}) \oplus (\bm{3},\bm{3})
$,
that composes Lorentz tensors like
\begin{equation}
\varphi^{\mu\nu}\bb{x} = A^{\mu\nu}\bb{x} + S^{\mu\nu}\bb{x} + \frac{1}{4}g^{\mu\nu} \Omega\bb{x},
\end{equation}
which correspond to a decomposition into smaller {\em irreps} related to the Poincar\'e classes quoted at \cite{extfields}, with $A^{\mu\nu} \equiv (\bm{1},\bm{3}) \oplus (\bm{3},\bm{1})$ totally anti-symmetric under $\mu\leftrightarrow \nu$, $S^{\mu\nu}\equiv (\bm{3},\bm{3})$ totally symmetric under $\mu\leftrightarrow \nu$, and $\Omega \equiv (\bm{1},\bm{1})$ transforming as a Lorentz scalar, multiplied by the metric tensor, $g^{\mu\nu}$.

As a matter of completeness, the above properties, as discussed in Ref.~\cite{extfields}, support the inclusion of interacting fields which also transform according to Poincar\'e symmetries described by the extended Poincar\'e group \cite{WuTung}, as they appear
in a full Dirac Hamiltonian like \cite{diraclike01,diraclike02}
\begin{eqnarray}
\label{E04T}
\hat{H}  &=& A^0\bb{\bm{x}}\,\hat{I}_4+ \hat{\beta}( m + \phi_S \bb{\bm{x}} ) + \hat{\bm{\alpha}} \cdot (\hat{\bm{p}} - \bm{A}\bb{\bm{x}}) + i \hat{\beta} \hat{\gamma}_5 \mu\bb{\bm{x}} - \hat{\gamma}_5 q\bb{\bm{x}} + \hat{\gamma}_5 \hat{\bm{\alpha}}\cdot\bm{W}\bb{\bm{x}} \nonumber \\
&+& i \hat{\bm{\gamma}} \cdot [ \zeta_a \bm{B}\bb{\bm{x}} + \kappa_a\, \bm{E}\bb{\bm{x}}  \,] + \hat{\gamma}_5 \hat{\bm{\gamma}}\cdot[\kappa_a\, \bm{B}\bb{\bm{x}}  - \zeta_a \bm{E}\bb{\bm{x}} \,],
\end{eqnarray}
where a fermion with mass $m$ and momentum $\bm{p}$ interacts with an external vector field with time component $A^0\bb{\bm{x}}$ and spatial components $\bm{A} \bb{\bm{x}}$, and is non-minimally coupled to magnetic and electric fields $\bm{B}\bb{\bm{x}}$ and $\bm{E}\bb{\bm{x}}$ (via $\kappa_a$ and  $\zeta_a$). The above Hamiltonian also admits the inclusion of pseudovector field interactions with time component $q\bb{\bm{x}}$, and spatial components $\bm{W}\bb{\bm{x}}$, besides both scalar and pseudoscalar field interactions through $\phi_S\bb{\bm{x}}$  and $\mu\bb{\bm{x}}$, respectively.
Algebraic strategies \cite{extfields} for obtaining the analytical expression for the matrix density of the associated eigenstates of the above Hamiltonian problem have been developed, however, they are out of the central scope of this paper.

\section{Bispinor entanglement under Lorentz boosts}

With the normalized bispinors from Eq.~(\ref{twoqubitspinor22}), one can construct a general quantum state of two-particles, $A$ and $B$, respectively with momentum (energy) $\bm{p}$ ($E_p$) and $\bm{q}$ ($E_q$), as a generic $M$-term normalized superposition,
\begin{eqnarray}
\label{generalstate}
\vert \, \Psi^{AB}\bb{\bm{p}, \bm{q}} \, \rangle &=&
\frac{1}{\sqrt{N}} \displaystyle \sum_{i=1}^M c_i \, \vert \, u_{s_i}\bb{\bm{p}} \rangle^A \otimes \vert \, u_{r_i}\bb{\bm{q}} \, \rangle^B,
\end{eqnarray}
with the normalization given by $\sum_{i=1}^M \vert c_i\vert^2 = N$, and where the subindex ``$_{\pm}$'' has been suppressed from the notation. Such two-particle states can be generated, for instance, in a QED elastic scattering process \cite{solano} \footnote{The choice of different momenta, $\{\bm{p}_i\}\neq \bm{p}$ for each particle state of the same vector subspace (either $A$ or $B$) introduces additional quantum correlations between spin and momemtum variables, turning the problem into a more complex and non-realistic one.}

As a matter of convenience, $u_{s_i}\bb{\bm{p}}$ (as well as $u_{r_i}\bb{\bm{q}}$) can be described by helicity eigenstates such that $\bm{e}_{p} \cdot \hat{\bm{\sigma}}^{(S)} \vert \chi_{s_i} \bb{\bm{p}} \rangle = (-1)^{s_i} \vert \chi_{s_i} \bb{\bm{p}}\rangle$ (where $\bm{e}_p = \bm{p}/\vert \bm{p} \vert$) can be factorized from Eq.~(\ref{twoqubitspinor22}) to set $u_{s_i}\bb{\bm{p}}$ a spin-parity separable state.
In terms of projected states $\vert z_\pm \rangle$, eigenstates of $\hat{\sigma}_z^{(S)}$, one can write
\begin{equation}
\vert \chi_{s_i}\bb{\bm{p}} \rangle = \frac{\hat{I}^{(S)}_2 + (-1)^{s_i} \bm{e}_{p} \cdot \hat{\bm{\sigma}}^{(S)}}{\sqrt{1 + \vert \bm{e}_{p} \cdot \bm{e}_z} \vert}\vert z_\pm \rangle,
\end{equation}
and, if $\bm{e}_{p}$ is in the $z$-direction, $\bm{e}_{p} \equiv \bm{e}_z$, one has $\vert \chi_1\bb{\bm{p}} \rangle = \vert z_+ \rangle$ and $\vert \chi_2\bb{\bm{p}} \rangle = \vert z_- \rangle$, such that, from now on, the labels $s_i$ (and also $r_i$), when they are set equal to $1$ and $2$, denote positive and negative helicity, respectively.

Under the above assumptions, the density matrix of the generic superposition from Eq.~(\ref{generalstate}) is written as
\begin{equation}
\label{2partDM}
\rho\bb{\bm{p}, \bm{q}} = \frac{1}{N} \displaystyle \sum_{i, j}^M c_i c_j^*\, \rho_{s_i  s_j}^{A}\bb{\bm{p}} \otimes \rho_{r_i  r_j}^{B}\bb{\bm{q}},
\end{equation}
where 
\begin{eqnarray}
\rho_{s_i  s_j}^{A}\bb{\bm{p}} &=&\left( \vert u_{s_i}\bb{\bm{p}} \rangle \langle u_{s_j}\bb{\bm{p}}\vert\right)^A \nonumber \\
&=&\frac{1}{2 E_{p}} \Bigg[ \left( E_p \delta_{s_i s_j} + m \delta_{s_i s_j+1} \right) \hat{I}^{(P)A}_2 + \left(E_p \delta_{s_i s_j+1} + m \delta_{s_i s_j} \right) \hat{\sigma}_z^{(P)A} + \nonumber \\
&&\qquad \qquad + \sqrt{E_{p}^2 - m^2}\left( (-1)^{s_j} \, \hat{\sigma}_+^{(P)A} + (-1)^{s_i} \, \hat{\sigma}_-^{(P)A} \right)\Bigg] \otimes \Xi_{s_i s_j}^{(S)A}\bb{\bm{p}}, 
\end{eqnarray}
where $\hat{\sigma}_{\pm} = \hat{\sigma}_x \pm i\hat{\sigma}_y$ and the factorized dependence on the momentum direction is implicitly given by
\begin{equation}
\Xi_{s_i s_j}^{(S)A}\bb{\bm{p}} = \left( \vert \chi_{s_i}\bb{\bm{p}} \rangle \langle \chi_{s_j}\bb{\bm{p}} \vert\right)^A,
\end{equation}
with a similar expression for $\rho_{r_i r_j}^{B}\bb{\bm{q}}$ by replacing $\{\bm{p}; s_{i(j)} \}$ by $\{\bm{q}; r_{i(j)}\}$ and $A$ by $B$. As each of the components of the state (\ref{2partDM}) is a two-qubit state, the joint state $\rho\bb{\bm{p}, \bm{q}}$ is a four-qubit state. Differently from the case where a system composed by two subsystems has the quantum entanglement supported by the Schmidt decomposition theorem, the classification and quantification of entanglement in the above constructed multipartite states is an open problem. Subsystems in a multipartite state can share entanglement in different non-equivalent ways, and the corresponding multipartite entanglement can be approached by different points of view. As the joint state (\ref{2partDM}) is a pure state, its corresponding multipartite entanglement can be computed through the Meyer-Wallach global measure of entanglement, $E_{G}[\rho]$, expressed in terms of the linear entropy, $E_{L}[\rho]$, as \cite{globalMW}
\begin{equation}
\label{globalent}
E_{G}[\rho] = \bar{E}[\rho^{\{\alpha_k\}}] = \frac{1}{4}\big[\, E_L[\rho^{(S)A}] + E_L[\rho^{(P)A}] + E_L[\rho^{(S)B}] + E_L[\rho^{(P)B}] \, \big],
\end{equation} 
with $$E_L[\rho] = \frac{d}{d-1}(1 - \mbox{Tr}[\rho^2]),$$ where $d$ is the dimension of the underlying Hilbert space in which $\rho$ acts, and the reduced density matrix of a given subsystem $\alpha_k$ is obtained by tracing out all the other subsystems $\rho^{\alpha_j} = \mbox{Tr}_{\{\alpha_k \} \neq \alpha_j} [\rho]$.
In the above problem, the subsystems considered correspond to spin and parity, $S$ and $P$, for particles $A$ and $B$, i.e. $\{\alpha_k\} \equiv \{(S)A,\,(S)B,\,(P)A,\,(P)B\}$. In particular, the more the subsystems of a given state are mixed, the more entanglement is encoded among them: the global measure, $E_{G}[\rho]$, captures a picture of the quantum correlations distributed among the four DoF's here involved..

The linear entropy of a reduced subsystem $\rho^{\alpha_k}$ of (\ref{generalstate}), which is a two-qubit state, is evaluated in terms of the components of its Bloch vector $a^{\alpha_k}_n = \mbox{Tr}[\hat{\sigma}_n^{\alpha_k} \rho^{\alpha_k}]$ as
\begin{equation}
\label{globalexpr0}
E_L[\rho^{\alpha_k}]= 1 - \sum_{n = \{x, \, y, \, z\}} (a^{\alpha_k}_n)^2,
\end{equation} 
and the global measure from Eq.~(\ref{globalent}) can be simplified into
\begin{equation}
\label{globalexpr}
E_{G}[\rho] =1 - \frac{1}{4} \, \displaystyle \sum_{\alpha = \{\alpha_k\}} \sum_{n = \{x, \, y, \, z\}} (a^{\alpha}_n)^2,
\end{equation}
with $ \{\alpha_k\}\equiv \{(S)A,\,(S)B,\,(P)A,\,(P)B\}$.
The Bloch vectors of the subsystems of $A$ are explicitly given by
\begin{eqnarray}
\label{blochvecs}
a^{(S)A}_n &=& \frac{1}{N}\displaystyle \sum_{i, j}^M\, c_i c_j^* \, \mathcal{M}_{r_i r_j}\bb{\bm{q}} \, \frac{1}{ E_p} ( \, E_p \delta_{s_i s_j} + m \delta_{s_i s_j+1} \,) \mbox{Tr}[\hat{\sigma}_n^{(S)A} \Xi_{s_i s_j}^{(S)A}\bb{\bm{p}}],
\end{eqnarray}
for the spin subsystem, 
\begin{eqnarray}
a^{(P)A}_x &=& \frac{1}{N} \displaystyle \sum_{i}^M (-1)^{s_i} \, \vert \, c_i \, \vert^2 \, \frac{\sqrt{E_p^2 - m^2}}{E_p}, \nonumber \\
a^{(P)A}_y &=& 0, \nonumber \\
a^{(P)A}_z &=& \frac{1}{N}\displaystyle \sum_{i, j}^M c_i c_j^* \, \mathcal{M}_{r_i r_j}\bb{\bm{q}} \, \mbox{Tr}[\Xi_{s_i s_j}^{(S)A}\bb{\bm{p}} ] \, \frac{1}{E_p} (\, E_p \delta_{s_i s_j+1} + m \delta_{s_i s_j} \,),
\end{eqnarray} 
for the parity subsystem, where 
\begin{eqnarray}
\label{coeff}
\mathcal{M}_{r_i r_j}\bb{\bm{q}} = \mbox{Tr}[\rho^{B}_{r_i r_j}\bb{\bm{q}}] = \frac{1}{E_q} ( \, E_q \delta_{r_i r_j} + m \delta_{r_i r_j+1} \,) \mbox{Tr}[\Xi_{r_i r_j}^{(S)B}\bb{\bm{q}}].
\end{eqnarray}
Analogous expressions for the Bloch vectors of the subsystems of $B$ are given by (\ref{blochvecs}) and (\ref{coeff}) with the replacement $\{\bm{p}; s_{i(j)} \}\leftrightarrow\{\bm{q}; r_{i(j)}\}$ and $A\leftrightarrow B$.

To evaluate the quantum entanglement encoded only by the spin DoF's in (\ref{2partDM}), one considers the spin-spin reduced density matrix
\footnotesize\begin{eqnarray}
\label{unboostspin}
\rho^{(S)A, (S)B}\bb{\bm{p}, \bm{q}} &=& \mbox{Tr}_{(P)A, (P)B} \left[\rho\bb{\bm{p}, \bm{q}}\right]  \\
&=& \frac{1}{N} \displaystyle \sum_{i,j}^M c_i c_j^* \, \frac{ \left( E_p \delta_{s_i s_j} + m \delta_{s_i s_j+1} \right) \, \left( E_q \delta_{r_i r_j} + m \delta_{r_i r_j+1} \right)}{E_p\,E_q}\, \Xi_{s_i s_j}^{(S)A}\bb{\bm{p}}\otimes \Xi_{r_i r_j}^{(S)B}\bb{\bm{q}},\quad\nonumber
\end{eqnarray}\normalsize
which is, in general, a mixed state.

Entanglement in mixed states cannot be evaluate in terms of the linear entropy, as a mixed subsystem does not imply into a joint entangled state for mixtures. Instead, the characterization of quantum entanglement, in this case, is given by the Peres separability criterion \cite{negativity01} which asserts that a bipartite state $\rho \in \mathcal{H}_{A} \otimes \mathcal{H}_B$ is separable iff the partial transpose density matrix with respect to the any of its subsystem, $\rho^{T_A}$, has only positive eigenvalues. With respect to a fixed basis on the composite Hilbert space $\{\vert \lambda_i \rangle \otimes \vert \nu_j \rangle \}$ (with $\vert \lambda_i \rangle \in \mathcal{H}_{A}$ and $\vert \nu_i \rangle \in \mathcal{H}_{B}$), the matrix elements of the partial transpose with respect to the $A$ subsystem $\rho^{T_{A}}$ are given by 
\begin{equation}
\langle \lambda_i \vert \otimes \langle \nu_j \vert (\,\rho \,)\,^{T_{A}} \vert \lambda_k \rangle \otimes \vert \nu_l \rangle = \langle \lambda_k \vert \otimes \langle \nu_j \vert \, \rho \, \vert \lambda_i \rangle \otimes \vert \nu_l \rangle,
\end{equation}
and in the light of the separability criterion, the negativity $\mathcal{N}[\rho]$ is defined as \cite{negativity02}
\begin{equation}
\label{negativity}
\mathcal{N}[\rho] = \displaystyle{\sum}_{i} \vert \lambda_i \vert - 1 ,
\end{equation}
where $\lambda_i$ are the eigenvalues of $\rho^{T_{A}}$. The spin-spin negativity of (\ref{2partDM}) $\mathcal{N} \big[ \rho^{(S)A, (S)B} \big]$ is then evaluated with the eigenvalues of the partial transpose of (\ref{unboostspin}) with respect to $(S)A$ as to return
\footnotesize\begin{eqnarray}
\label{unboostparttranspose}
\big ( \, \rho^{(S)A,(S)B)} \, \big )^{T_{A}}\bb{\bm{p}, \bm{q}} &=& \big( \,\mbox{Tr}_{(P)A, (P)B} \left[\rho\bb{\bm{p}, \bm{q}}\right] \, \big)^{T_{A}} \\
&=& \frac{1}{N} \displaystyle \sum_{i,j}^M c_i c_j^* \, \frac{ \left( E_p \delta_{s_i s_j} + m \delta_{s_i s_j+1} \right) \, \left( E_q \delta_{r_i r_j} + m \delta_{r_i r_j+1} \right)}{E_p\,E_q}\, \Xi_{s_j s_i}^{(S)A}\bb{\bm{p}}\otimes \Xi_{r_i r_j}^{(S)B}\bb{\bm{q}},\quad\nonumber
\end{eqnarray}\normalsize
where the subtle change with respect to (\ref{unboostspin}) is in the subindex of $\Xi^{(S)A}$.

\section{Covariance of the Dirac equation and the effects of Lorentz boosts}

Once the global and the spin-spin entanglement of the general superposition (\ref{generalstate}) are characterized by Eqs.~(\ref{globalexpr}) and (\ref{blochvecs}), and the spin-spin negativity is evaluated through the eigenvalues of Eq.~(\ref{unboostparttranspose}), one can describe how the Lorentz boosts do affect such quantum correlations. First, one notices that the covariant form of the Dirac equation
\begin{equation}
(\hat{\gamma}_\mu p^\mu -m \hat{I}_4) \psi\bb{x} = 0,
\end{equation}
where $\hat{\gamma}_0 = \hat{\beta}$ and $\gamma_{\mu} = (\gamma_{0}, \hat{\bm{\gamma}})$ with $\hat{\bm{\gamma}} = \hat{\beta} \hat{\bm{\alpha}}$, transforms under a Lorentz boost, $x^\mu \rightarrow x^{\mu\prime} = \Lambda^\mu_{\, \, \, \nu} x^\nu$, as
\begin{eqnarray}
(\hat{\gamma}^\mu p_\mu -m \hat{I}_4) \psi\bb{x} = 0 \, \, \rightarrow \, \, ((\hat{\gamma}^\prime)^\mu p^\prime_\mu -m \hat{I}_4) \psi^\prime(x^\prime) = 0,
\end{eqnarray}
and its solution, $\psi \bb{x}$, transforms as
\begin{equation}
\label{boooo}
\psi^\prime(x^\prime) = \hat{S}[\, \Lambda \,] \psi(\Lambda^{-1} x^\prime),
\end{equation}
where $\hat{S}[\, \Lambda \,]$ corresponds to the transformation in the bispinor space representation (cf. $\hat{D} [\Lambda]$ from (\ref{spinstates})). Lorentz boosts, $\Lambda (\omega)$, can be parameterized in terms of components in the vector representation of the $SO(1,3)$ as
$[\Lambda (\omega)]_{ij} = \delta_{ij} + (\cosh (\omega) - 1)\,n_i \, n_j$, $[\Lambda (\omega)]_{i0} = [\Lambda(\omega)]_{0i} = \sinh{(\omega)} \,n_i$, and $[\Lambda (\omega)]_{00} = \cosh{(\omega)}$, where $\omega=\mbox{arccosh}(\sqrt{1-v^2})$ is the (dimensionless) boost rapidity, $v$ is the reference frame velocity (between $\mathcal{S}$ and $\mathcal{S}^{\prime}$) and $n_i$ are the space components of the boost direction, $\bm{n}$, with $\bm{n}\cdot\bm{n}=1$. In the bispinor space representation, $\hat{S}[\Lambda(\omega)]$, reads
\begin{equation}
\label{boostrep}
\hat{S}[\Lambda(\omega)]= \cosh{\left( \frac{\omega}{2} \right)} \hat{I}_4 - \sinh{\left( \frac{\omega}{2} \right)} \bm{n} \cdot \hat{\bm{\alpha}},
\end{equation}
which is a non-unitary operator. 
By following the above introduced two-qubit prescription, the boost operator (\ref{boostrep}) can be expressed in the form of
\begin{equation}
\label{twoqubitboost}
\hat{S}[\Lambda(\omega)]=\cosh{\left( \frac{\omega}{2} \right)} \hat{I}_2^{(P)}\otimes \hat{I}_2^{(S)} - \sinh{\left( \frac{\omega}{2} \right)} \bm{n} \cdot( \, \hat{\sigma}_x^{(P)} \otimes \hat{\bm{\sigma}}^{(S)} \,),
\end{equation}
from which one can evaluate the effects of boosts in parity and spin subsystems. For instance, keeping the covariant notation for the quadrimomentum, $p$, the density matrix of a single helicity bispinor with quantum number $s$ transforms under boosts as
\begin{equation}
\rho_s\bb{p} \rightarrow \rho_s^\prime\bb{p^\prime} = \frac{1}{\cosh(\omega)}\hat{S}[\Lambda(\omega)] \, \rho_s(\Lambda^{-1} p^\prime) \, \hat{S}^\dagger[\Lambda(\omega)],
\end{equation}
where the term $(\cosh{(\omega)})^{-1}$ was included as to keep the normalization of the spinor, and (\ref{twoqubitboost}) can be used to describe the transformation law of the subsystem described by the spin density, $\rho^{(S)}_s\bb{p} = \mbox{Tr}_{(P)}[ \rho_s\bb{p}]$, as
\begin{eqnarray}
\rho^{(S)}_s\bb{p} \rightarrow \rho^{\prime(S)}_s\bb{p^\prime} = \frac{1}{\cosh{(\omega)}} \Big[ \cosh^2{\left(\frac{\omega}{2} \right)} \rho^{(S)}_s\bb{p} + \sinh^2{\left(\frac{\omega}{2} \right)} (\bm{n} \cdot \hat{\bm{\sigma}}) \rho^{(S)}_s\bb{p} (\bm{n} \cdot \hat{\bm{\sigma}}) \nonumber \\
\qquad \qquad\qquad\qquad\qquad\qquad\qquad - (-1)^s \sinh{(\omega)} \frac{E_p - m}{E_p} \{ \bm{n} \cdot \hat{\bm{\sigma}}, \rho^{(S)}_s\bb{p} \}\Big],
\end{eqnarray}
where $\{\,\,,\,\,\}$ denotes anti-commutators, and which, in the limit $E_p - m \simeq E_p$, can be subtly simplified as to give a transformation law in the form of $\rho^{\prime(S)}_s\bb{p^\prime} =\hat{O} \, \rho^{(S)}_s\bb{p} \, \hat{O}^\dagger$, where $\hat{O}$ is the unitary operator
\begin{equation}
\hat{O} = \frac{1}{\sqrt{\cosh{(\omega)}}} \left[ \cosh{\left(\frac{\omega}{2} \right)} \hat{I}_2 - \sinh{\left(\frac{\omega}{2} \right)} (\bm{n} \cdot \hat{\bm{\sigma}}) \left({\bm{e}}_{p} \cdot \hat{\bm{\sigma}} \right) \right].
\end{equation}
In fact, such transformation under a Lorentz boost is the same as that one obtained for states belonging to the {\em irrep} $(+, \frac{1}{2})$ of the Poincar\'{e} group, which can be recast in terms of a momentum dependent rotation and which is the basis of several results in relativistic quantum information\footnote{The non-unitarity of $\hat{S}[\Lambda(\omega)]$ has also additional implications for the definition of spin operators in the context of relativistic quantum mechanics \cite{spins}. Apart from the usual Pauli spin operator $\propto\hat{\bm{\Sigma}} = \hat{I}^{(P)}_2 \otimes \hat{\bm{\sigma}}^{(S)}$, other spin operators were also proposed in the literature. For example, the Fouldy-Wouthuysen (FW) spin operator \cite{FW} was used in the context of transformation properties of Dirac bispinors as to define a covariant spin reduced density matrix \cite{celeri, bi-spinorFW}, and states constructed with FW eigenstates were then used in describing transformation properties of spin entropy as well as spin-spin Bell's inequality under Lorentz boosts.}. 

Considering the generic two-particle state (\ref{2partDM}) in a reference frame $\mathcal{S}$, the transformed density matrix describing the state in an inertial frame $\mathcal{S}^{\prime}$, related to $\mathcal{S}$ by a Lorentz boost, $\Lambda$, is given by
\begin{eqnarray}
\rho\bb{\bm{p}, \bm{q}} \rightarrow \rho^\prime\bb{\bm{p}^\prime, \bm{q}^\prime} &=& \frac{1}{\nu} \big( \, \hat{S}^{A}[\Lambda] \otimes \hat{S}^{B}[\Lambda] \,\big) \,\rho\bb{\bm{p}, \bm{q}} \, \big( \, (\hat{S}^{A}[\Lambda])^\dagger \otimes (\hat{S}^{B}[\Lambda])^\dagger \,\big) \nonumber \\
&=&\frac{1}{\nu} \sum_{i,j}^M c_i c_j^* \, \varrho_{s_i s_j}^{ A} \bb{\bm{p}}\otimes \varrho_{r_i r_j}^{B}\bb{\bm{q}},
\end{eqnarray}
where $\nu = Tr\left[\left( \, \hat{S}^{A}[\Lambda] \otimes \hat{S}^{B}[\Lambda] \right)^2\,\rho\bb{\bm{p}, \bm{q}}\right]$ and the transformed term $ \varrho_{s_i s_j}^{A}\bb{\bm{p}}$ reads
\begin{eqnarray}
 \varrho_{s_i s_j}^{ A}\bb{\bm{p}} &=& \cosh^2{\left( \frac{\omega}{2} \right)} \rho_{s_i \, s_j}^{A}\bb{\bm{p}} - \frac{\sinh(\omega)}{2} \{ \rho_{s_i s_j}^{A}\bb{\bm{p}}, \, (\hat{\sigma}_x^{(P)A} \otimes \bm{n} \cdot \hat{\bm{\sigma}}^{(S)A}) \, \} \nonumber \\ &&\qquad\qquad\qquad+ \sinh^2{\left( \frac{\omega}{2} \right)} (\hat{\sigma}_x^{(P)A} \otimes \bm{n} \cdot \hat{\bm{\sigma}}^{(S)A}) \, \rho_{s_i s_j}^{A}\bb{\bm{p}} \, (\hat{\sigma}_x^{(P)A} \otimes \bm{n} \cdot \hat{\bm{\sigma}}^{(S)A}), 
\end{eqnarray}
with an analogous expression for $\varrho_{r_i r_j}^{B}\bb{\bm{q}}$. The difference between the global entanglement in $\mathcal{S}^{\prime}$ and $\mathcal{S}$,
\begin{equation}
\Delta E_G = E_G[\rho^\prime\bb{\bm{p}^\prime, \bm{q}^\prime}] - E_G[\rho\bb{\bm{p}, \bm{q}}],
\end{equation}
is evaluated through Eqs.~(\ref{globalexpr}) and (\ref{blochvecs}) replaced by transformed Bloch vectors, now renamed by $a\to \mathcal{A}$, which are given by
\footnotesize\begin{eqnarray}
\label{transfblochspin}
{\mathcal{A}}^{(S)A}_{k} &=& \frac{1}{\nu}\displaystyle \sum_{i,j}^M c_i c_j^* \, \mu_{r_i r_j} \, \frac{1}{E_p}\Bigg[ \, \mbox{Tr}[\hat{\sigma}_k^{(S)A} \Xi_{s_i s_j}^{(S)A}] \cosh^2{\left( \frac{\omega}{2} \right)} ( \, E_p \delta_{s_i s_j} + m \delta_{s_i s_j +1} \,) \nonumber \\
&&\qquad-2 \, (-1)^{s_i} n_k \, \mbox{Tr}[\Xi_{s_i s_j}^{(S)A}] \, \sinh(\omega) \sqrt{E_p^2 - m^2} \, \delta_{s_i s_j}  \\
&&\qquad\qquad\qquad+ \mbox{Tr}[\hat{\sigma}_k^{(S)A}\, (\bm{n} \cdot \hat{\bm{\sigma}}^{(S)A})\, \Xi_{s_i s_j}^{(S)A} \, (\bm{n} \cdot \hat{\bm{\sigma}}^{(S)A})] \sinh^2{\left( \frac{\omega}{2} \right)} ( \, E_p \delta_{s_i s_j} + m \delta_{s_i s_j +1} \,) \Bigg]\nonumber,
\end{eqnarray}\normalsize
for the spin reduced subsystem of $A$,  and
\begin{eqnarray}
\label{transfblochpar}
{\mathcal{A}_x}^{(P)A} &=& \frac{1}{\nu}\displaystyle \sum_{i,j}^M c_i c_j^* \, \mu_{r_i r_j} \, \frac{1}{E_p} \, \Bigg[(-1)^{s_i} \mbox{Tr}\big[\Xi_{s_i s_j}^{(S)A} \big] \cosh(\omega) \sqrt{E_p^2 - m^2} \delta_{s_i s_j} \nonumber \\
&&\qquad -\sinh(\omega) \, \mbox{Tr}\big[\,(\bm{n} \cdot \hat{\bm{\sigma}}^{(S)A} )\Xi_{s_i s_j}^{(S)A} \, \big] \left( E_p \delta_{s_i s_j} + m \delta_{s_i s_j +1} \right) \Bigg], \nonumber \\
{\mathcal{A}_x}^{(P)A} &=& \frac{1}{\nu}\displaystyle \sum_{i,j}^M c_i c_j^* \, \mu_{r_i r_j} \, \frac{E_p \delta_{s_i s_j+1} + m \delta_{s_i s_j}}{E_p} \mbox{Tr}\big[\Xi_{s_i s_j}^{(S)A} \big],
\end{eqnarray}
for the parity reduced subsystem, where
\begin{eqnarray}
\label{transfcoeff}
 \mu_{r_i r_j} &=& \mbox{Tr}[\varrho_{r_i r_j}^{ B}] \nonumber \\
&=& \frac{1}{E_p}\big[ \, \cosh{(\omega)}(\, E_q \delta_{r_i r_j} + m \delta_{r_i r_j+1}\,) \mbox{Tr}[\Xi_{r_i r_j}^{(S)B}] \nonumber \\
&&\qquad\qquad - (-1)^{r_{i}} \, \sinh{(\omega)} \delta_{r_i r_j} \sqrt{E_q^2 - m^2} \, \mbox{Tr}[\bm{n} \cdot \hat{\bm{\sigma}}^{(S)B} \, \Xi_{r_i r_j}^{(S)B}]\, \big],
\end{eqnarray}
and, in all the above expressions, the explicit dependence on $\bm{p}$ and $\bm{q}$ has been suppressed from the notation. Through the above expressions, again, the Bloch vector for the subsystems of $B$ can be obtained with the replacement $\{\bm{p}; s_{i(j)} \}\leftrightarrow\{\bm{q}; r_{i(j)}\}$ and $A\leftrightarrow B$ into Eqs.~(\ref{transfblochspin})-(\ref{transfcoeff}). For any boost one also has ${\mathcal{A}_y}^{(P)A} = {\mathcal{A}_y}^{(P)B}=0$.

The effects of the boost on the spin-spin entanglement, on the other hand, are described by the change on the negativity
\begin{equation}
\Delta \mathcal{N}^{(S)A,(S)B} = \mathcal{N}[\varrho^{(S)A,(S)B}] -\mathcal{N}[\rho^{(S)A,(S)B}] ,
\end{equation}
with the transformed spin-spin density matrix given by
\begin{eqnarray}
\label{spintransformed}
\varrho^{(S)A,(S)B} =\frac{1}{\nu} \displaystyle \sum_{i,j}^M c_i c_j^* \, \varrho_{s_i s_j}^{ (S)A} \otimes \varrho_{r_i r_j}^{(S)B},
\end{eqnarray}
where
\begin{eqnarray}
\varrho_{s_i s_j}^{ (S)A} &=&\cosh^2{\left( \frac{\omega}{2} \right)}\, \frac{E_p \delta_{s_i s_j} + m \delta_{s_i s_j +1}}{E_p} \, \Xi_{s_i s_j}^{(S)A} \nonumber \\
&&\qquad - (-1)^{s_i} \frac{ \sinh(\omega) }{2} \frac{\sqrt{E_p^2 - m^2}}{E_p} \delta_{s_i s_j} \{ \Xi_{s_i s_j}^{(S)A}, \bm{n} \cdot \hat{\bm{\sigma}}^{(S)A} \} \nonumber \\
&&\qquad\qquad +\sinh^2{\left( \frac{\omega}{2} \right)} \,\frac{E_p \delta_{s_i s_j} + m \delta_{s_i s_j +1}}{E_p} (\bm{n} \cdot \hat{\bm{\sigma}}^{(S)A}) \, \Xi_{s_i s_j}^{(S)A}\,( \bm{n} \cdot \hat{\bm{\sigma}}^{(S)A}),
\end{eqnarray}
with a corresponding expression for $\varrho_{r_i r_j}^{(S)B}$. From the above expression one concludes that if the boost is performed in a direction $\bm{n}$ such that $ \{ \Xi_{s_i s_j}^{(S)A}, \bm{n} \cdot \hat{\bm{\sigma}}^{(S)A} \}= \{ \Xi_{r_i r_j}^{(S)B}, \bm{n} \cdot \hat{\bm{\sigma}}^{(S)B} \} = 0$, then the spin reduced density matrix (\ref{spintransformed}) is invariant.

\subsection{Entanglement for an overall class of anti-symmetric states}

The above framework describes quantitatively the changes on multipartite quantum correlations, as quantified by $E_G$, and on spin-spin entanglement induced by Lorentz boosts acting on a generic superposition of two-particle helicity bispinors, as quantified by $\mathcal{N}$. As the nature of fermionic particles requires anti-symmetric wave functions, states that are given by the anti-symmetric superpositions have to be considered in the form of
\begin{equation}
\label{antisymmetric}
\vert \Psi^{odd}_{sr}\bb{\bm{p}, \bm{q}} \rangle=\frac{\vert u_s\bb{\bm{p}} \rangle^{A} \otimes \vert u_r\bb{\bm{q}} \rangle^{B} - \vert u_r\bb{\bm{q}} \rangle^{A} \otimes \vert u_s\bb{\bm{p}} \rangle^{B}}{\sqrt{2}}.
\end{equation}

Talking about Dirac particles like electrons, quarks, neutrinos, etc, some of the above configurations are very difficult to be produced phenomenologically. Thus, only some examples shall be considered in the following, from the less to more relevant ones.

\begin{figure}
\centering
\includegraphics[width= 6 cm]{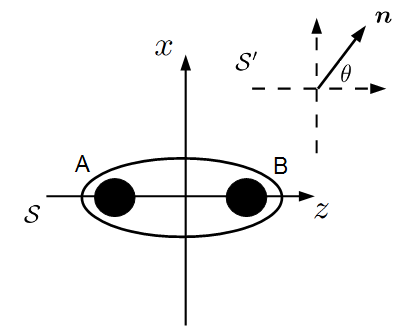}
\renewcommand{\baselinestretch}{1.0}
\caption{Schematic representation of the boost scenario. The joint state of $A$ and $B$ is described by the anti-symmetric superposition of bispinors (\ref{antisymmetric}) with momenta, in $\mathcal{S}$, $\bm{p} = - \bm{q} = p \bm{e}_z$. A Lorentz boost is performed as to describe the joint state on a frame $\mathcal{S}^{\prime}$ moving with respect to $\mathcal{S}$ in the $\bm{n} = \sin{(\theta)} \bm{e}_x + \cos{(\theta)} \bm{e}_z$ direction with rapidity $\omega$.}
\label{Scheme}
\end{figure}

At the reference frame $\mathcal{S}$ with $\bm{p} = - \bm{q}$, the center of momentum frame, positive and negative helicity eigenstates are given by  
\begin{eqnarray}
\label{hel}
\vert \chi_1 \bb{\bm{p}} \rangle = \vert \chi_2 \bb{\bm{q}} \rangle = \vert z_+ \rangle, \nonumber \\
\vert \chi_2 \bb{\bm{p}} \rangle = \vert \chi_1 \bb{\bm{q}} \rangle = \vert z_- \rangle,
\end{eqnarray}
and, in the unboosted frame $\mathcal{S}$, the states are also eigenstates of the Pauli spin operator, $\sigma_z$. It is sufficient to consider the boost with direction $\bm{n}$ in a plane defined by the unitary vectors, $\bm{e}_z$ and $\bm{e}_x$, with $\bm{n} = \sin{(\theta)} \bm{e}_x + \cos{(\theta)} \bm{e}_z$ as pictorially depicted in Fig.~\ref{Scheme}.
 
\begin{figure}
\centering
\includegraphics[width= 16 cm]{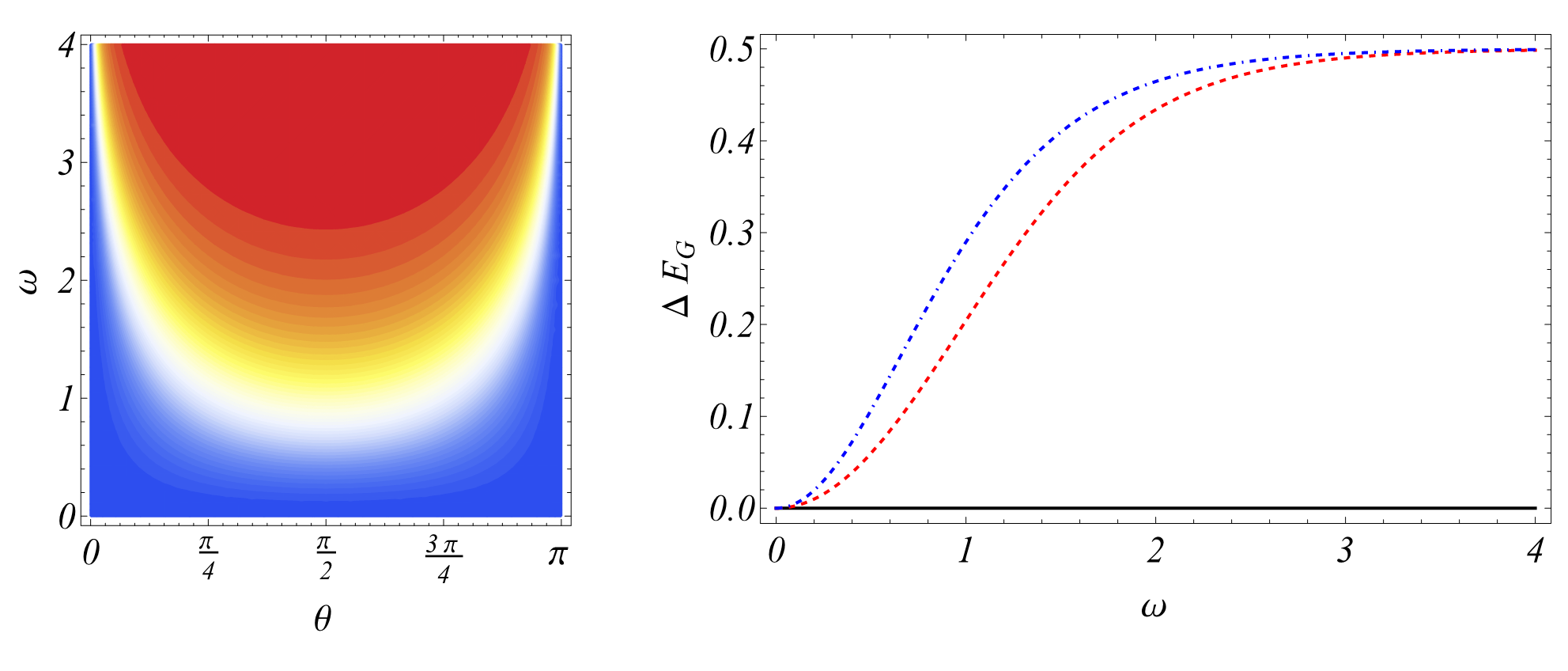}
\renewcommand{\baselinestretch}{1.0}
\caption{Variation of the global entanglement for the state (\ref{state1}) as function of the boost rapidity $\omega$ (dimensionless) and of the boost direction angle $\theta$ in radians (left plot) and as function of the boost rapidity for $\theta = 0$ (black solid line), $\pi/4$ (red dashed line) and $\pi/2$ (blue dotdashed line). The initial rapidity of the states in the unboosted frame is $\omega_0 = \mbox{arccosh}(E_p/m) = 1$. The global entanglement encoded in the DoF's of such pair of bispinor always increases due to the boost when the transformation is in a direction not parallel to the momenta $\bm{p}$ and $\bm{q}$ with respect to $\mathcal{S}$. In the limit of high speed boosts, $E_G$ has its maximum value $1$, as for the unboosted state $E_G [\rho]=1/2$. Among this correlations, no spin-spin entanglement is present as in any frame $\mathcal{N}^{(S)A,(S)B}[\psi_1] = 0$.}
\label{Graph01}
\end{figure}

By adapting the notation to the simplifications from Eq.~(\ref{hel}), one has the anti-symmetric state given by\begin{equation}
\label{state1}
\vert \psi_1 \rangle = \frac{\vert u_1\bb{\bm{p}} \rangle^{A} \otimes \vert u_2\bb{\bm{q}} \rangle^{B} - \vert u_2\bb{\bm{q}} \rangle^{A} \otimes \vert u_1\bb{\bm{p}} \rangle^{B}}{\sqrt{2}},
\end{equation}
in a superposition of helicities which, however, is spin-spin separable. Since $\Xi_{s \, r}^{(S)A}= \Xi_{s \, r}^{(S)B}=\vert z_+ \rangle \langle z_+ \vert$ for all $s$ and $r$, the transformed spin-spin density matrix Eq. (\ref{spintransformed}) is invariant under partial transposition with respect to any of its subsystems, and thus a Lorentz boost does not create spin-spin entanglement.  Nevertheless, the global entanglement $E_G$ is not invariant, as depicted in Fig.~\ref{Graph01} which shows $\Delta E_G$ as function of the boost rapidity $\omega$ and of the boost angle $\theta$. Of course, this is because $\vert \psi_1 \rangle$ mixes different momentum eigenstates, in a kind of artificial and unrealistic physical composition of particles $A$ and $B$. Boosts parallel to the momenta in $\mathcal{S}$ does not increase the amount of global entanglement in the state, although for any non-parallel boosts the global entanglement increases due to an increasing in both parity and spin reduced entropies, which are essentially constrained by the dependence on the momentum components. It tends to the maximum value ($\sim 1$) for high-speed boosts.

\begin{figure}
\centering
\includegraphics[width= 16 cm]{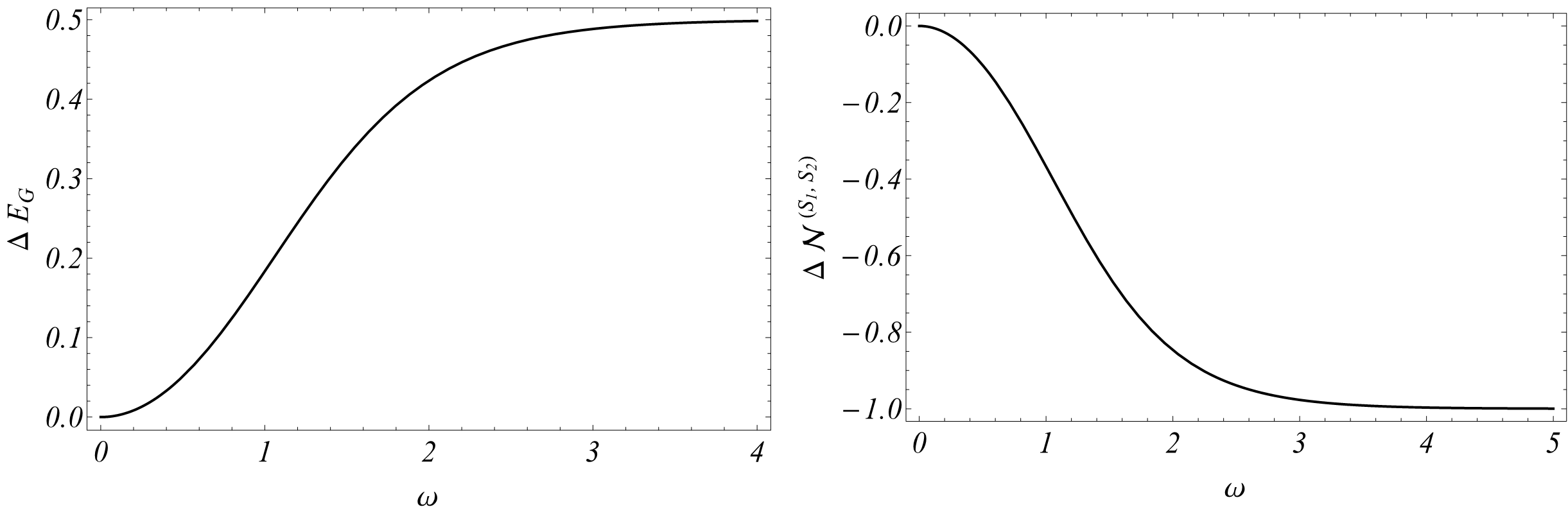}
\renewcommand{\baselinestretch}{1.0}
\caption{Variation of the global entanglement (left plot) and spin-spin entanglement (right plot) as function of the rapidity $\omega$ (dimensionless) for the state (\ref{state2}). The transformation of correlations in this state is independent from the boost angle $\theta$ and, although the global entanglement increases due to the boost with a behavior similar to the one of Fig.~(\ref{Graph01}), the spin-spin entanglement is degraded. In the limit of high-speed boosts, $S_{AB}$ is separable from all the other DoF's and spin-spin entanglement vanishes: the global entanglement is encoded by the parity DoF's.}
\label{Graph02}
\end{figure}

Otherwise, a maximally entangled spin-spin state in $\mathcal{S}$ can be constructed through an anti-symmetric superposition between positive helicities
\begin{equation}
\label{state2B}
\vert \psi_2 \rangle = \frac{\vert u_1\bb{\bm{p}} \rangle^{A} \otimes \vert u_1\bb{\bm{q}} \rangle^{B} - \vert u_1\bb{\bm{q}} \rangle^{A} \otimes \vert u_1\bb{\bm{p}} \rangle^{B}}{\sqrt{2}},
\end{equation}
which, according to the correspondence from (\ref{hel}), indeed can be recast as
\begin{equation}
\label{state2}
\vert \psi_2 \rangle = \frac{\vert u_1\bb{\bm{p}} \rangle^{A} \otimes \vert u_1\bb{\bm{q}} \rangle^{B} - \vert u_2\bb{\bm{p}} \rangle^{A} \otimes \vert u_2\bb{\bm{q}} \rangle^{B}}{\sqrt{2}},
\end{equation}
which corresponds to a much more realistic configuration, for which particles in the subspace $A$ and $B$ have well defined momenta, $\bm{p}$ and $\bm{q}$, respectively, in agreement with the construction from the previous section.
Fig.~\ref{Graph02} depicts the variation of the global and the spin-spin entanglement of $\vert \psi_2 \rangle$ as function of the boost rapidity $\omega$. In this case, the variation of entanglement is independent of the boost angle and, as for the state from Eq.~(\ref{state1}), the global entanglement increases under Lorentz boosts. On the other hand, spin-spin entanglement is degraded by the boost transformation and for high speed boosts the spin-spin state is completely separable.

\begin{figure}
\centering
\includegraphics[width= 16 cm]{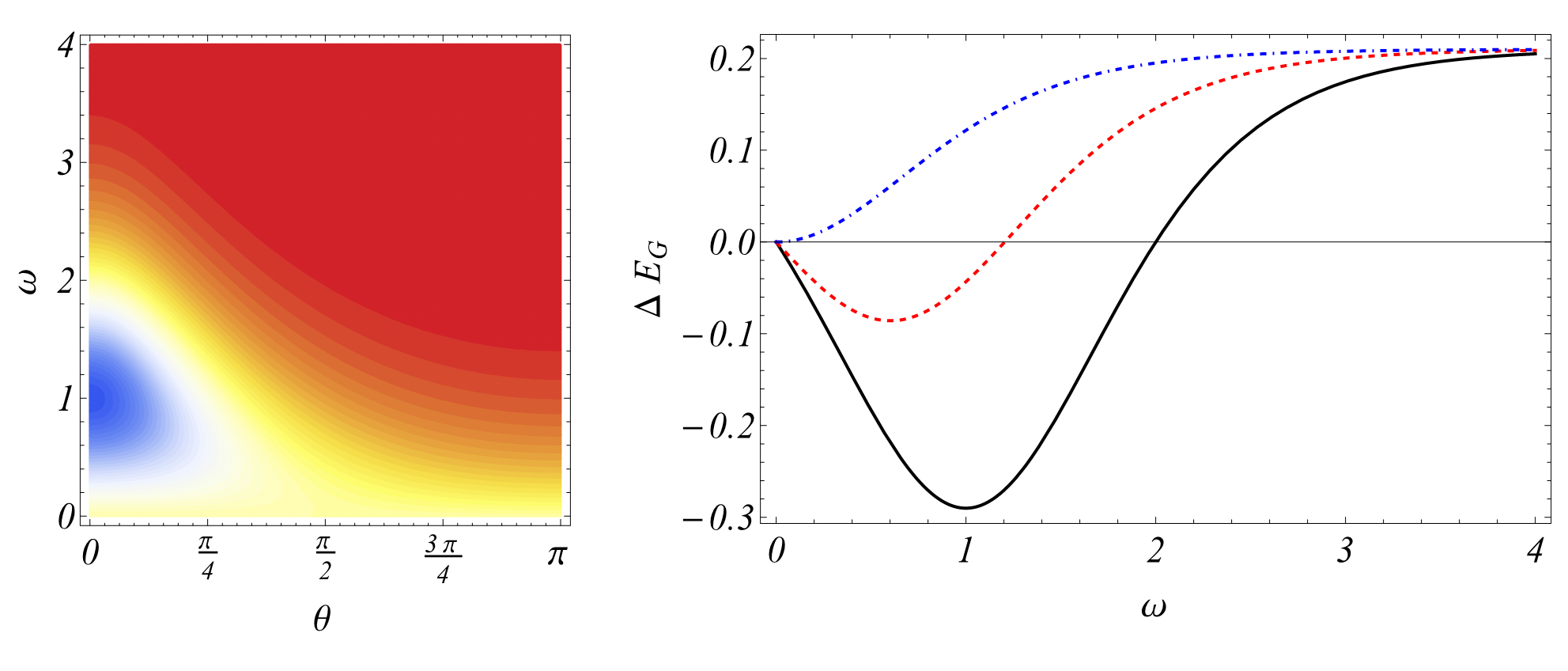}
\renewcommand{\baselinestretch}{1.0}
\caption{Variation of the global entanglement under Lorentz boost for the state (\ref{state3}) as function of the boost rapidity $\omega$ (dimensionless) and of the boost direction angle $\theta$ in radians (left plot). The curves of the right plot are in correspondence with those of Fig.~\ref{Graph01}. Different from Figs.~\ref{Graph02} and \ref{Graph03}, global entanglement exhibit a non-monotonous behavior for $\theta < \pi/2$. For a parallel boost $\theta = 0$ (solid curve) global entanglement reaches its minimum for $\omega=1$, which corresponds to the reference frame in which the bispinors are at rest: all quantum correlations correspond to only spin-spin entanglement.}
\label{Graph03}
\end{figure}

\begin{figure}[b]
\centering
\includegraphics[width= 16 cm]{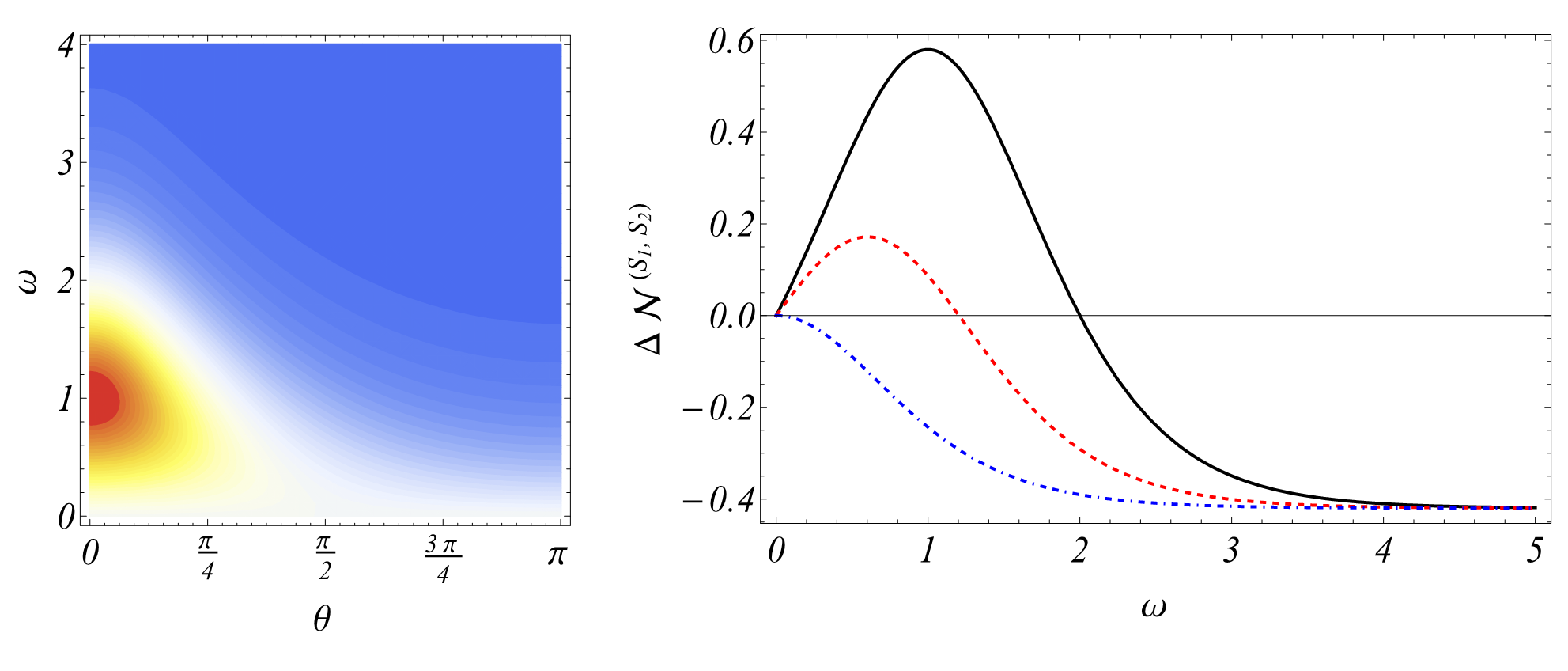}
\renewcommand{\baselinestretch}{1.0}
\caption{Variation of spin-spin entanglement under Lorentz boost for the state (\ref{state3}) as function of the boost rapidity $\omega$ (dimensionless) and of the boost direction angle $\theta$ in radians (left plot) and in function of the boost rapidity for boosts direction in correspondence with those of  Figs.~\ref{Graph01, Graph03}. The behavior of quantum correlations encoded between the spins degrees of freedom is complementary to the one exhibit by the global entanglement of Fig.~\ref{Graph03}. For $\theta < \pi/2$ the behavior is non-monotonous with an local maximum corresponding to the rest frame of the bispinors, and a complete degradation resulting from a high-speed boost.}
\label{Graph04}
\end{figure}

A third anti-symmetric configuration is given by
\begin{equation}
\label{state3}
\vert \psi_3 \rangle = \frac{\vert u_1\bb{\bm{p}} \rangle^{A} \otimes \vert u_2\bb{\bm{p}} \rangle^{B} -\vert u_2\bb{\bm{p}} \rangle^{A} \otimes \vert u_1\bb{\bm{p}} \rangle^{B}}{\sqrt{2}},
\end{equation}
which describes a two-particle helicity superposition moving in the $\bm{e}_z$ direction where both particles have the same momenta. This case is phenomenologically interesting because $\Delta v = 0$ is a kinematical Lorentz invariant.
Two electrons in a common rest frame will have $\Delta v = 0$ for any relativistic boost. In this case, the spin-spin entanglement depends on the momentum $p$ even in the unboosted frame. Differently from the preliminary examples, both global entanglement, depicted in Fig.~\ref{Graph03}, and spin-spin entanglement, depict in Fig.~\ref{Graph04}, exhibit a non-monotonous behavior under Lorentz boosts. In particular, for a boost parallel to the momentum $\bm{p}$ with rapidity equals to $\mbox{arccosh}(\, E_p/m \,)$, the global entanglement is minimum, as this frame corresponds to the common rest frame of the particles where there is only spin-spin entanglement. For a high speed boost, the entanglement shared between the DoF's of the state is enhanced, although the spin-spin entanglement, as in the case of state (\ref{state2}), is completely degraded.

It is worth to mention that, although the global measure from Eq.~(\ref{globalent}) was considered, four-qubit state entanglement can be computed through another global measure of entanglement defined in a similar fashion of (\ref{globalent}), but with linear entropies of the reduced subsystems of two qubits. This quantity is calculated with terms of the form $\mbox{Tr}[\hat{\sigma}^{\alpha_k}_i \hat{\sigma}^{\beta_l}_j \rho^{\{\alpha_k; \beta_l\}}]$ and contains, in addition to the information encoded in $E_G$ (\ref{globalent}), also correlations between pairs of the subsystems \cite{rigolin}. Nevertheless, in the case of the anti-symmetric states considered here, the behavior of this quantity is qualitatively similar to the behavior depicted in Figs.~\ref{Graph02} - \ref{Graph03} and add no information about the variation of quantum entanglement encoded by bispinors under Lorentz boosts. Other point of view of multipartite entanglement is provided by considering the geometry of the composite Hilbert space, and by studying distances between a given multipartite state and the set of the so-called $K$-separable states \cite{MassimoEnt}. In this case, the quantification of multipartite entanglement can capture more information about different multipartite components that contribute to the total amount of quantum correlations in a given state, requiring an extremization process. This more complete picture of multipartite entanglement for the two spinors states considered here is postponed to future investigations.

\subsection{Transformation of entanglement in chiral states}

Superpositions of eigenstates of the chiral operator $\hat{\gamma}_5 = \hat{\sigma}_x^{(P)} \otimes, \hat{I}^{(S)}$ defined in terms of the free bispinors $u_s\bb{\bm{p}}$ as
\begin{equation}
u^{f}_s\bb{\bm{p}} = \frac{\hat{I} + (-1)^f \hat{\gamma}_5}{2} u_s\bb{\bm{p}},
\end{equation}
with $f = 0, 1$, can also be investigated in the above context.
Differently from the helicity, the chirality is a Lorentz invariant given that the chiral and the boost operator commute, i.e. $[\hat{\gamma}_5, \hat{S}[\Lambda(\omega)]\,] = 0$.
However, for massive particles, it is not a dynamical conserved quantity as $[\hat{\gamma}_5, \hat{H}] \neq 0$ \cite{Alex001,Alex002}.
This invariance property has implications for the transformation laws of quantum entanglement encoded by superpositions of chiral states
\begin{equation}
\label{chiral01}
\psi^{Chiral}\bb{\bm{p},\bm{q}} = \frac{1}{N}\displaystyle \sum_{i}^M c_i \, \vert u^{f_i} _{s_i}\bb{\bm{p}}\rangle^A \otimes \vert u^{g_i}_{r_i}\bb{\bm{q}}\rangle^B,
\end{equation}
where $f_i$ is the chirality of the bispinor $\vert u^{f_i} _{s_i}\bb{\bm{p}}\rangle^A$ and $g_i$ is the chirality of $\vert u^{g_i}_{r_i}\bb{\bm{q}}\rangle^A$. Chiral states constructed through projection of helicity states can be written in the simplified form of
\begin{equation}
\vert u^{f}_s\bb{\bm{p}} \rangle = \vert f \rangle \otimes \vert \chi_s \bb{\bm{p}} \rangle
\end{equation}
where $\vert f \rangle = (\vert z_+ \rangle + (-1)^f \vert z_- \rangle)/2$ are the eigenstates of $\hat{\sigma}_x$ operator, and thus the density matrix of (\ref{chiral01}) reads
\begin{equation}
\label{chiral02}
\rho_{Chiral} = \frac{1}{N}\displaystyle \sum_{i,j}^M c_i c_j^* \,( \vert f_i \rangle \langle f_j \vert )^{A}\otimes \Xi^{(S)A}_{s_i s_j} \otimes ( \vert g_i \rangle \langle g_j \vert )^{B} \otimes \Xi^{(S)B}_{r_i r_j},
\end{equation}
where, again, the explicit dependence on momenta has been suppressed.
Since the chiral eigenstates are invariant under boosts, the density matrix (\ref{chiral02}) transforms as
\begin{equation} 
\rho_{Chiral}^\prime = \frac{1}{N}\displaystyle \sum_{i,j}^M c_i c_j^* \,( \vert f_i \rangle \langle f_j \vert )^{A}\otimes \Xi^{\prime \, (S)A}_{s_i s_j} \otimes ( \vert g_i \rangle \langle g_j \vert )^{B} \otimes \Xi^{\prime \, (S)B}_{r_i r_j}, 
\end{equation}
where
$\Xi^{\prime \,(S)A}_{i\, j} = \hat{\mathcal{O}}_{f_i} \, \Xi^{(S)A}_{s_i s_j} \, \hat{\mathcal{O}}_{f_j}$, 
with
\begin{eqnarray}
\hat{\mathcal{O}}_{f_i} &=& \cosh{\left( \frac{\omega}{2}\right)} \hat{I} - (-1)^{(f_i)} \sinh{\left( \frac{\omega}{2}\right)} \bm{n} \cdot \hat{\bm{\sigma}},
\end{eqnarray}
and changes on the global entanglement are exclusively due to changes on the spin terms $\Xi_{s_i \, s_j}^{(S)A}$. A particular situation is for $f_i = f$ and $g_i = g$ for which
$\Xi^{\prime \, (S)A}_{i\, j} = \hat{\mathcal{O}}_{f} \, \Xi^{(S)A}_{s_i s_j} \, \hat{\mathcal{O}}_{f}
$, and 
\begin{eqnarray}
\rho_{Chiral}^\prime &=& \frac{1}{N}\displaystyle \sum_{i,j}^M c_i c_j^* \,( \vert f \rangle \langle f \vert )^{A}\otimes \Xi^{\prime \, (S)A}_{i\, j} \otimes ( \vert g \rangle \langle g \vert )^{B} \otimes \Xi^{\prime \, (S)B}_{i\, j}, \nonumber
\end{eqnarray}
which exhibits an invariant quantum correlation when anti-symmetric states as from Eqs.~(\ref{state2})-(\ref{state3}) are considered. In fact, the chiral states
\begin{equation}
\vert \, \psi_{2(3)}^{Chiral} \, \rangle = \left( \frac{\hat{I} + (-1)^f \hat{\gamma}_5}{2} \right)^{A} \otimes \, \left( \frac{\hat{I} + (-1)^g \hat{\gamma}_5}{2} \right)^{B}\vert \psi_{2 (3)} \rangle,
\end{equation}
with $f, g = 0,1$, are such that, for a boost direction given by $\bm{n} = (\sin(\theta), \, 0, \, \cos(\theta) )$, one has
$\rho_{2 (3)}^{Chiral} \rightarrow \rho_{2 (3)}^{\prime \, Chiral} = \rho_{2 (3)}^{Chiral},
$ and the states are completely Lorentz invariant.

\section{Conclusions}

The relativistic transformation properties of quantum entanglement have been on the focus of many recent investigations, with a special interest in describing how the spin-spin entanglement does change under Lorentz boosts. Although the setup usually adopted to describe transformation properties of quantum entanglement has given some interesting insights into the physics of relativistic quantum information, when massive charged fermions are considered as the physical carriers of spin one-half, a more complete description of the problem is required. The physical particles, such as electrons, muons, etc., are described by QED including, apart from the usual Poincar\'{e} symmetry, also invariance under parity transformation. This last symmetry operation exchange two {\em irreps} of the Poincar\'{e} group, and a proper formulation is given in terms of {\em irreps} of the so called complete Lorentz group. The states of the particles are then described by four component objects, the Dirac bispinors, which satisfy the Dirac equation.

In this paper we have described how Lorentz boosts do affect quantum entanglement shared among the DoF's of a pair of bispinorial particles in a generic framework. As each of the bispinors is supported by a $SU(2) \otimes SU(2)$ structure associated with the spin and intrinsic parity, the corresponding multipartite entanglement was quantified by means of the Meyer-Wallach global measure of entanglement, given in terms of the linear entropies of each subsystem. Additionally, since the reduced spin state is mixed, the spin-spin entanglement was quantified through the appropriate negativity. By means of the $SU(2) \otimes SU(2)$ decomposition of the boost operator, $\hat{S}[\Lambda]$, the transformation laws for the Bloch vectors (and for the reduced spin density matrix) of each subsystem were recovered for a generic state, setting a framework to describe changes on both global and spin-spin entanglements.

In order to specialize our results we have considered the action of Lorentz boosts in three different anti-symmetric states. First we considered a spin-spin separable state in which the particles are moving in opposite directions in the unboosted frame. In such scenario, Lorentz boosts cannot create spin-spin entanglement and the global entanglement monotonously increase as a function of the boost rapidity. The second anti-symmetric state considered here describes particles with opposite momenta and maximal spin-spin entanglement. As in the first case, the global entanglement increases as consequence of the boost, although a degradation of spin-spin entanglement is induced by the frame transformation. The last specific case consists of a pair of particles with same momentum and spin-spin entanglement, exhibiting a non-monotonous behavior of both global and spin-spin entanglement under Lorentz boost. Finally, we addressed the effects of Lorentz boosts on chiral states, which exhibit some subtle invariance properties. In particular, the density matrices obtained through projections of the anti-symmetric states on definite chiral states are completely invariant under boosts.

The general formalism developed through this paper sets the framework for some future developments including the computation of quantum entanglement among particles involved in scattering processes \cite{scattering}. It may also be useful in the aim of a field theoretical description of relativistic entanglement. Finally, given that some low energy systems, such as trapped ions and graphene, emulate the Dirac equation dynamics \cite{diraclike02}, interactions in such systems can be engendered as to reproduce the effects of Lorentz transformations in feasible manipulable platforms which can work as simulating platforms for high energy physics measurements.

{\em Acknowledgments - The work of AEB is supported by the Brazilian Agencies FAPESP (grant 2017/02294-2) and CNPq (grant 300831/2016-1). The work of VASVB is supported by the Brazilian Agency CAPES (grant 88881.132389/2016-1).}

\end{document}